\documentclass[aps,prb,twocolumn,showpacs,superscriptaddress]{revtex4-1}
\usepackage{epsfig}
\usepackage[T1]{fontenc}
\usepackage[latin9]{inputenc}
\usepackage{color}
\usepackage{amsmath, amsthm, amssymb}
\usepackage[latin9]{inputenc}
\usepackage{hyperref}
\usepackage{graphicx}
\usepackage{subfigure}

\newcommand{\bcen}{\begin{center}}
\newcommand{\ecen}{\end{center}}
\newcommand{\bec}{}
\newcommand{\btab}{\begin{tabular}}
\newcommand{\etab}{\end{tabular}}
\newcommand{\bdes}{\begin{description}}
\newcommand{\edes}{\end{description}}

\newcommand{\beq}{\begin{equation}}
\newcommand{\eeq}{\end{equation}}
\newcommand{\bea}{\begin{eqnarray}}
\newcommand{\eea}{\end{eqnarray}}

\newcommand{\bary}{\begin{array}}
\newcommand{\eary}{\end{array}}
\newcommand{\benum}{\begin{enumerate}}
\newcommand{\eenum}{\end{enumerate}}
\newcommand{\exenum}{\begin{enumerate}[label=(\alph{*}),leftmargin=8mm]}
\newcommand{\bitem}{\begin{itemize}}
\newcommand{\eitem}{\end{itemize}}

\newcommand{\bk} { \mbox{\boldmath $k$}}

\newcommand{\bK} { \mbox{\boldmath $K$}}

\def\c2{ CuO$_2~$}

\def\he4{${\rm {}^4He}$~}

\begin{document}


\title{Lifshitz Transition in the Two Dimensional Hubbard Model}


\author{K.-S. Chen}
\email{kchen5@lsu.edu}
\affiliation{Department of Physics and Astronomy, Louisiana State University, Baton Rouge, LA 70803, USA}
\author{Z. Y. Meng}
\affiliation{Department of Physics and Astronomy, Louisiana State University, Baton Rouge, LA 70803, USA}
\affiliation{Center for Computation and Technology, Louisiana State University, Baton Rouge, LA 70803, USA}
\author{T. Pruschke}
\affiliation{Department of Physics, University of G\"ottingen, D-37077 G\"ottingen, Germany}
\author{J. Moreno}
 \affiliation{Department of Physics and Astronomy, Louisiana State University, Baton Rouge, LA 70803, USA}
 \affiliation{Center for Computation and Technology, Louisiana State University, Baton Rouge, LA 70803, USA}
 \author{M. Jarrell}
 \affiliation{Department of Physics and Astronomy, Louisiana State University, Baton Rouge, LA 70803, USA}
 \affiliation{Center for Computation and Technology, Louisiana State University, Baton Rouge, LA 70803, USA}

\date{\today}

\begin{abstract}

Using large-scale dynamical cluster quantum Monte Carlo simulations, we study the Lifshitz transition of the two dimensional Hubbard model with next-nearest-neighbor hopping ($t'$), chemical potential and temperature as control parameters. At $t'\le0$, we identify a line of Lifshitz transition points associated with a change of the Fermi surface topology at zero temperature. In the overdoped region, the Fermi surface is complete and electron-like; across the Lifshitz transition, the Fermi surface becomes hole-like and develops a pseudogap. 
At (or very close to) the Lifshitz transition points, a van Hove singularity in the density of states crosses the Fermi level. The van Hove singularity occurs at finite doping due to correlation effects, and becomes more singular when $t'$ becomes more negative. The resulting temperature dependence on the bare $d$-wave pairing susceptibility close to the Lifshitz points is significantly different from that found in the traditional van Hove scenarios. Such unambiguous numerical observation of the Lifshitz transition at $t'\le0$ extends our understanding of the quantum critical region in the phase diagram, and shines lights on future investigations of the nature of the quantum critical point in the two dimensional Hubbard model.
\end{abstract}


\maketitle

\section{Introduction}

The physical properties of high-\textit{$T_{c}$} cuprate superconductors are extremely sensitive 
to doping.  Experiments, such as the angle-resolved photoemission spectroscopy~\cite{Damascelli03} 
(ARPES) and quantum oscillation measurements~\cite{Vignolle08}, have clearly demonstrated a 
Fermi surface reconstruction as the doping concentration is varied. In the overdoped region, ARPES and quantum oscillation studies revealed a large Fermi surface that can be well captured by band theory. On the other hand, once the doping is reduced, the shape and size of Fermi surface change and photoemission data indicate that the Fermi surface breaks up into "Fermi arcs"\cite{Norman98}. More recent studies further indicate that the 
"Fermi arcs" are actually part of closed hole pockets~\cite{YangHB08,YangHB11}. The formation of 
small Fermi pockets, as the doping is reduced from overdoped to underdoped region, has also been
substantiated by quantum oscillation measurements~\cite{LeBoeuf07,Sebastian10}. 

From a theoretical point of view, the change of the Fermi surface topology from a large surface to small
"arcs" or "pockets" resembles a Lifshitz transition.
Although proposed by Lifshitz in noninteracting Fermion systems decades ago~\cite{Lifshitz60}, the Lifshitz transition has only recently begun to be considered as a quantum phase transition in strongly correlated electron systems~\cite{Yamaji06,Misawa07,Yamaji2007,Imada2010,Okamoto10,Imada2011,Bercx12}.
It has been used to explain experimental data in high-\textit{$T_{c}$} cuprate superconductors~\cite{Ovchinnikov09,Norman10} or heavy fermion systems \cite{Hackl11}.

Using large-scale dynamical cluster quantum Monte Carlo simulations~\cite{Maier05}, a series of recent numerical works~\cite{Marcridin06,Vidhyadhiraja09,Khatami10,YangSX11,Galanakis11,Chen11} mapped out the phase diagram of the two dimensional Hubbard model near the quantum critical filling. Particularly, the superconducting dome in the proximity to the quantum critical doping has been identified~\cite{YangSX11}. At positive $t'$, there is a first-order phase separation transition occurring at finite temperature. The two phases being separated are an incompressible Mott liquid and a compressible Mott gas; these two phases are adiabatically connected to the pseudogap and the Fermi liquid states at $t'=0$. The first-order line of coexistence terminates at a second order point where the charge susceptibility diverges~\cite{Khatami10,Galanakis11}. As $t'\rightarrow0$, this critical point extrapolates continuously to zero temperature and thus becomes the quantum critical point (QCP) underneath the superconducting dome. Above the QCP, a V-shaped quantum critical region, characterized as a marginal Fermi liquid~\cite{Varma96,Varma99} with linear resistivity, separates the pseudogap and the Fermi liquid phases. Furthermore, when the next-nearest-neighbor hopping becomes negative, $t'<0$, there is indication of the Fermi surface topology changes at zero temperature, and the filling at which such changes occurs is an extension of the QCP at $t'=0$ to negative $t'$\cite{Chen11}.

Other numerical works also found similar feature of the Fermi surface reconstruction at $t'\le 0$ by varying the hole doping concentration from overdoped towards half-filling. These include results from both dynamical cluster approximation~\cite{Gull09,Philipp09,Lin10,Gull10} and the cellular dynamical mean-field theory~\cite{Kotliar01,Sakai09,Liebsch09,Sordi11,Sakai10,Sakai12}, although the finite temperature critical point in the hole-doped side of 
the phase diagram found in Ref [35] is inconsistent with other works and might suffer from the finite size effects 
of a small four-site cluster.


Following such evidence, and in light of viewing the Fermi surface reconstruction as a Lifshitz transition~\cite{Yamaji06,Misawa07,Yamaji2007,Ovchinnikov09,Norman10,Imada2011}, we performed systematic numerical studies 
on the quantum critical phase diagram of the two dimensional Hubbard model at various $t'$, with special attention to the region of $t'<0$, which is relevant for the hole-doped cuprates. We find at $t'/t \le 0$, as the doping concentration varies from the overdoped to the underdoped regime, the Fermi surface changes its topology from electron-like with complete Fermi surface to hole-like with pseudogap at the anti-nodal direction. Such a topological transition in the Fermi surface is a Lifshitz transition. It is furthermore 
concomitant with a van Hove singularity in the density of states crossing the Fermi level at a doping which occurs very close to (if not at) the quantum critical point.

The van Hove singularity crossing and the concomitant quantum critical point occur at finite doping due to correlation effects even when $t'=0$. Interestingly, we find the quantum critical phenomena prevail for negative $t'$, and the van Hove singularity defines a line of quantum critical points which extends from the QCP at $t'=0$ to 
higher doping for $t'<0$. The temperature dependence of correlation effects close to the van Hove singularities, and its influence on quantities like the quasiparticle fraction and the pairing polarization are very different from those found in the traditional van Hove scenarios~\cite{Hirsch86b,Markiewicz97}. The QCP and the van Hove singularity have great impact on the conditions for pairing. At $t'<0$, we find an enhanced temperature dependence of the bare $d$-wave pairing susceptibility above the QCP that cannot be captured by the conventional BCS logarithmic divergence or the log-squared divergence found at a static van Hove singularity~\cite{Hirsch86b}.

This paper is organized as follows. Section II outlines the model and the methods used in this study: the dynamic cluster approximation (DCA) with weak-coupling continuous-time quantum Monte Carlo (CTQMC) as its cluster solver. Section III and IV contain our numerical results and discussion, beginning with the spectral function and dispersion at various doping concentrations and values of $t'/t$, and followed by a detailed account of other single-particle properties such as density of states and quasiparticle fraction, across the Lifshitz transition. We then provide results on the unique temperature dependence of the bare d-wave pairing susceptibility. We use a schematic quantum critical phase diagram of the model which summarizes our numerical results. We end with conclusions and an overview of open questions in Section V.


\section{Formalism}
\label{sec:formalism}

In this work, we look for direct evidence of a Lifshitz transition in the spectral function of 
the two-dimensional Hubbard model  
\begin{equation}
H=\sum_{\bk\sigma}\epsilon_{\bk}^{0}c_{{\bk}\sigma}^{\dagger}
c_{{\bk}\sigma}^{\phantom{\dagger}}+U\sum_{i}n_{{i}\uparrow}n_{{i}\downarrow},
\label{eq:hubbard}
\end{equation}
where $c_{{\bk}\sigma}^{\dagger}(c_{{\bk}\sigma})$ is the creation (annihilation) operator for 
electrons with wavevector ${\bk}$ and spin $\sigma$,  $n_{i\sigma} =c_{i\sigma}^{\dagger}c_{i\sigma}$ 
is the number operator, and the bare dispersion is given by
\begin{equation}
\epsilon_{\bk}^{0}=-2t\left(\cos k_{x}+\cos k_{y}\right)-4t'\left(\cos k_{x}\cos k_{y}-1\right)\, ,
\label{StdDisp}
\end{equation}
with $t$ and $t'$ being the hopping amplitude between nearest and the next-nearest-neighbor sites 
respectively, and $U$ the on-site Coulomb repulsion.

We employ the DCA~\cite{Hettler98, Hettler00} with weak-coupling CTQMC~\cite{Rubtsov05} as a cluster solver. The DCA is a cluster 
mean-field theory that maps the original lattice onto a periodic cluster of size $N_c=L_c^D$ ($D$ is the
dimensionality) embedded in 
a self-consistently determined host. It treats the spatial short-ranged correlations (up to $L_c$ 
inside a cluster) explicitly while approximating the long-ranged correlations with a mean-field. In 
this work we choose a square cluster with $N_c=16$. The six independent momentum patches are centered 
at $\Gamma=(0,0)$, $M=(\pi,\pi)$, $X=(\pi,0)$, $(\pi/2,\pi/2)$, $(\pi,\pi/2)$, and $(\pi/2,0)$. We set the 
energy scale to $4t=1$, choose the interaction strength at $U=6t$ and study inverse temperatures up to $\beta=58/4t$. 
The temporal correlations, essential for quantum criticality, are treated explicitly by the weak-coupling CTQMC 
solver for all cluster sizes. The solver expands the Coulomb interaction diagrammatically and samples in time continuously. Different from the Hirsch-Fye algorithm~\cite{Hirsch86a,Jarrell01}  it therefore  has no Trotter error. 

We obtain high-quality estimates of the cluster self-energy $\Sigma(\bK,\omega)$ by employing the maximum entropy 
analytical continuation~\cite{Jarrell96} (MEM) directly to the Matsubara-frequency self energies calculated from the 
DCA-CTQMC~\cite{Wang09,Chen11,Fuchs2011}. 
To perform MEM on the self-energy the non-Hartree part of  $\Sigma(\bK,i\omega_n)$ must be 
normalize by $U^2 \chi_{\sigma,\sigma}$, where $\chi_{\sigma,\sigma} = 
\left\langle n_\sigma n_\sigma\right\rangle -\left\langle n_\sigma \right\rangle^2 = n_\sigma(1-n_\sigma)$ 
is the local polarization of a single spin species $\sigma$. The normalized spectrum of 
the self-energy acts as a probability distribution:
\begin{equation}
{\frac{\Sigma(\bK,i\omega_n) - \Sigma_H}{U^2 \chi_{\sigma,\sigma}} }
= 
\int d\omega {\frac{\sigma(\bK,\omega)}{i\omega_n - \omega}},
\end{equation}
where $\displaystyle \sigma(\bK,\omega) = -\frac{1}{\pi} \Sigma^{\prime\prime}(\bK,\omega)/U^2 \chi_{\sigma,\sigma}$,
$\int d\omega {\sigma(\bK,\omega)} = 1$, using $\chi_{\sigma,\sigma}$ obtained from the Monte Carlo process. 


To obtain the lattice self energy, $\Sigma(\bk,\omega)$, we interpolate the cluster self 
energy, $\Sigma(\bK,\omega)$. From the lattice self energy we can get the lattice single-particle spectral function, $A(\bk,\omega)$. An alternative way has been suggested by Stanescu et al.~\cite{Stanescu06,Sakai09,Sakai10,Sakai12}, where the cluster cumulant, $M(\bK,\omega)=1/(\omega+\mu-\Sigma(\bK,\omega))$, has been used for the interpolation to lattice quantities.
Although the spectral functions produced by the two interpolation schemes have differences in the underdoped region, we found the two methods give the same quantum critical filling $n_c$~\cite{footnote12}.

\begin{figure*}[ht!]
\centerline{
\includegraphics[width=3.5in]{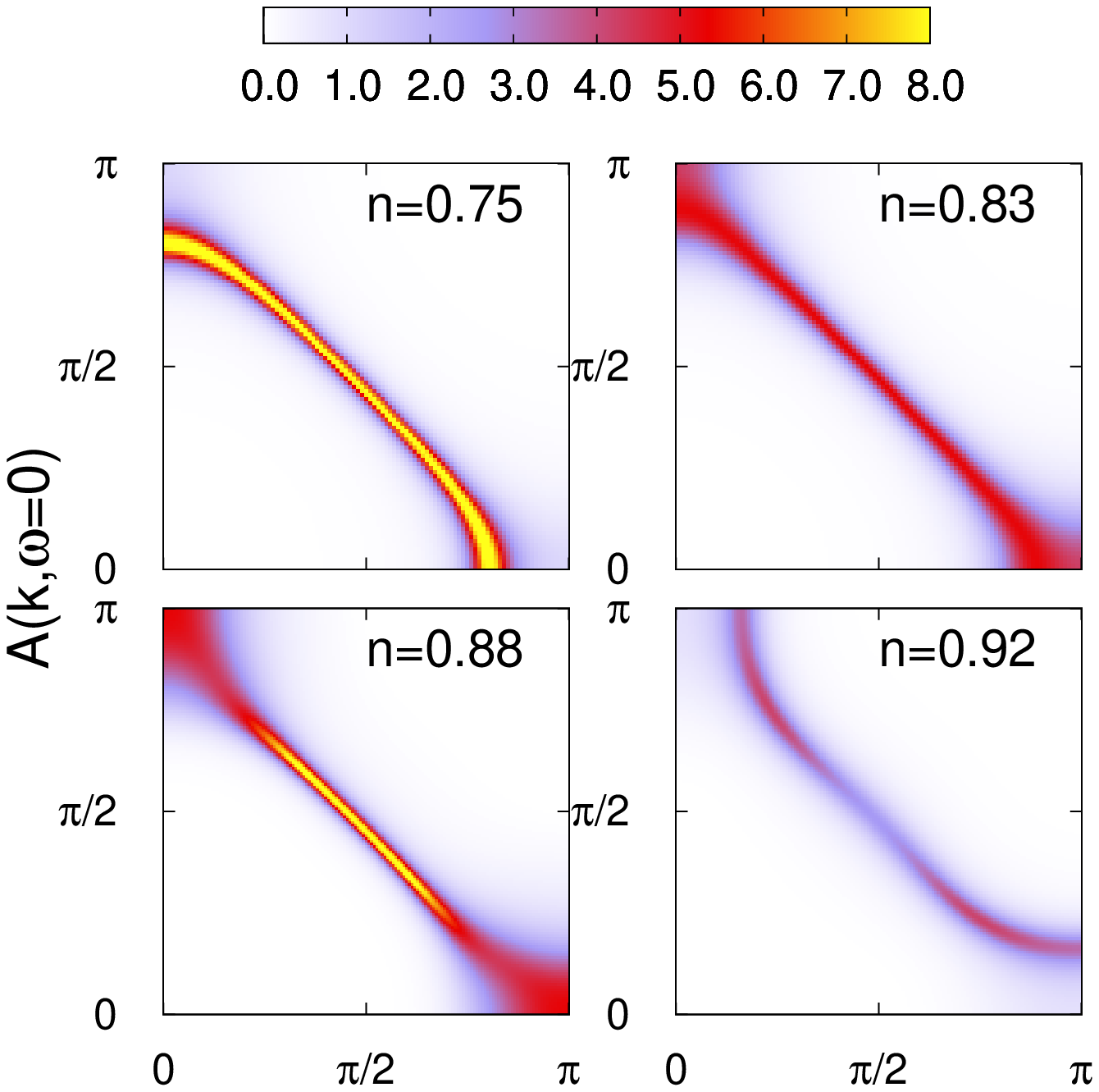}
\includegraphics[width=3.5in]{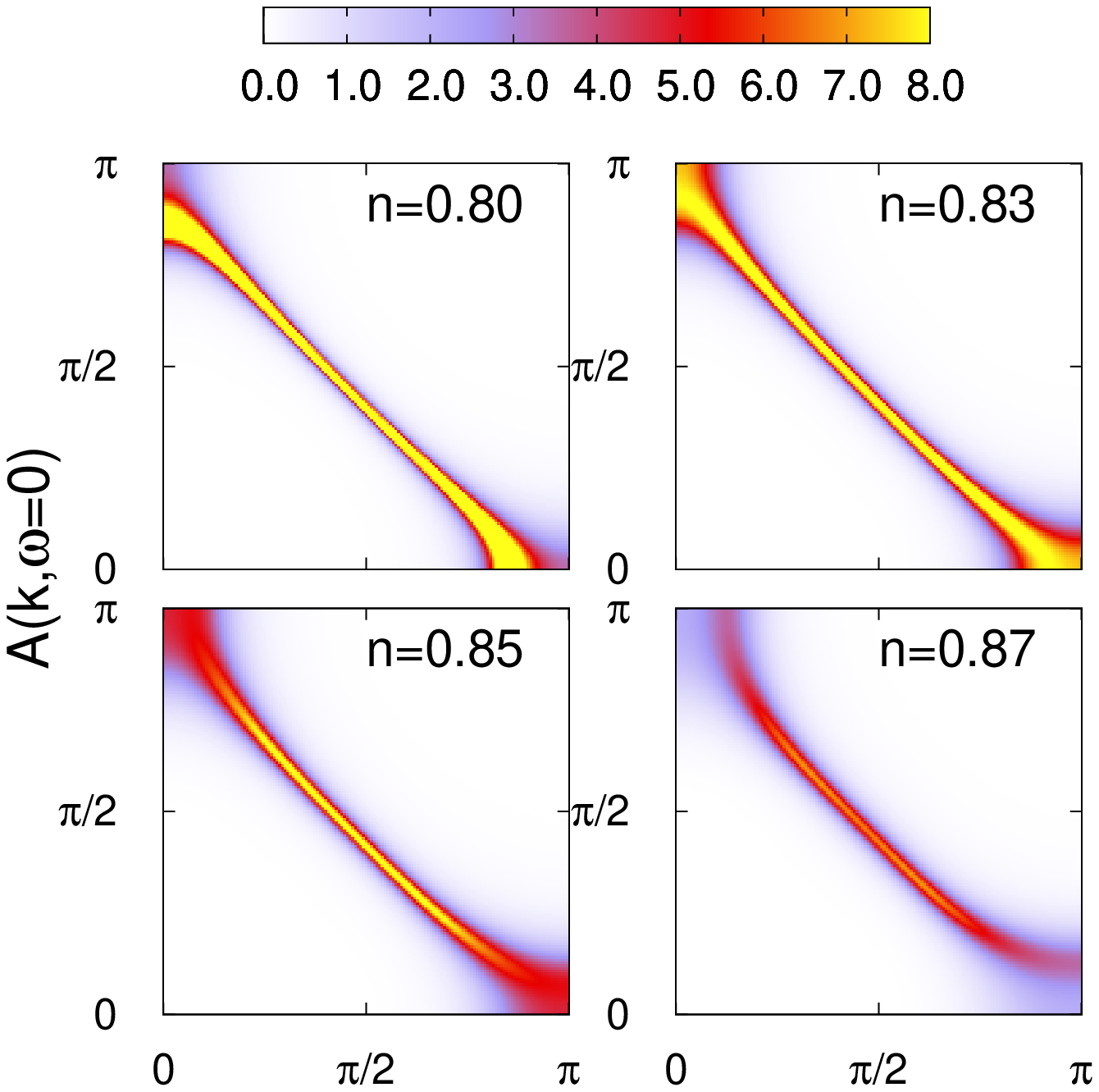}
}
\caption{(color online) Zero frequency spectra $A(\bk,\omega=0)$ for $t'/t=0$ (left) and $t'/t=-0.1$ (right) as a function 
of filling $n$. As $n$ increases towards half-filling, the system undergoes a Lifshitz 
transition where the Fermi surface changes topology from electron-like to hole-like.  
(Left panel)  The Lifshitz transition occurs at $n_c\approx0.88$, where the van Hove singularity crosses the Fermi level 
(as shown in Figure \ref{fig:DOS}).  (Right panel) The Lifshitz transition occurs at $n_c\approx0.835$. Note that when $t'/t=-0.1$, the Lifshitz transition happens at a smaller filling  than for the $t'=0$ case. 
In either case, as $n \to 1$, the system enters the pseudogap phase with vanishing quasiparticle 
weight.}
\label{fig:FalseColortp0tpm01}
\end{figure*}

\section{Results}
\label{sec:results}
Figure \ref{fig:FalseColortp0tpm01} shows the ${\bf k}$-resolved spectral function evaluated at the Fermi level, $A(\bk,\omega=0)$, for $t'/t=0$ and $t'/t = - 0.1$. At $t'=0$ we observe a change of the Fermi surface topology from electron-like to hole-like as the filling increases from $n=0.75$ towards half-filling.  
This change of the topology of the Fermi surface or Lifshitz transition  occurs at the same filling where the van Hove singularity crosses the Fermi level~\cite{Markiewicz97,Markiewicz12}. The corresponding filling is the critical filling, $n_{c}$, of the Lifshitz transition. 
In the case $t'=0$ we obtain $n_c\approx 0.88$.  At $t'/t=-0.1$ the Fermi surface changes from electron-like to hole-like and then disappears as $n$ varies from $0.8$ towards half filling. Following the same criterion, the Lifshitz transition 
occurs at $n_c\approx0.835$.  Finally, for $t'/t=-0.2$, we find $n_c\approx 0.77$ (not shown). 
For all $t'/t\le 0$ we studied, the system enters the pseudogap phase once the filling is larger than the corresponding $n_{c}$. In the case of $n=0.92$ (left panel of Fig.~\ref{fig:FalseColortp0tpm01}) and $n=0.87$ (right panel of 
Fig.~\ref{fig:FalseColortp0tpm01}) the collapses of the single-particle spectral weight along the antinodal direction can be clearly seen. 

\begin{figure}[h!]
\centering{}
\includegraphics[width=3.3in]{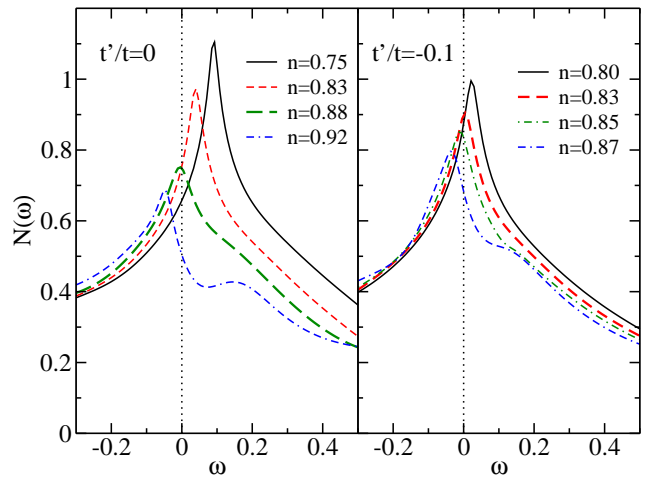}
\caption{(color online) The single-particle density of states for the parameters shown in Figure \ref{fig:FalseColortp0tpm01}. 
(Left panel) $t'/t=0$ with $n=0.75, 0.83, 0.88$ and $0.92$. (Right panel) $t'/t=-0.1$ with $n=0.80, 0.83, 0.85$ and $0.87$. 
The position of the van Hove singularity (peak in the density of states) shifts from positive frequency to negative frequency as the filling moves towards half-filling. 
The quantum critical point is at $n_c\approx0.88$ for $t'/t=0$, and $n_c\approx0.835$ for $t'/t=-0.1$.}
\label{fig:DOS}
\end{figure}

Figure \ref{fig:DOS} shows density of states for $t'/t=0$ (left panel) and $t'/t=-0.1$ (right panel) with the same set 
of fillings of those in Fig.~\ref{fig:FalseColortp0tpm01}. For $t'=0$, the van Hove singularity shifts from positive to negative frequency as the filling increases from 0.75 towards 1.0. The van Hove singularity is located at the Fermi level for the quantum critical filling $n_c\approx 0.88$. These results are consistent with our previous observations~\cite{Khatami10, Chen11}. After passing through the critical filling, 
a pseudogap in the density of states is formed, as it is displayed by  the valley at $\omega\approx0.05$ for $n=0.92$, $t'/t=0$. This parameter regime can thus be identified as the pseudogap region in the phase diagram. For $t'/t=-0.1$, the results in the right panel of Fig.~\ref{fig:DOS} show a quantum critical filling $n_c\approx 0.835$. Note, the van Hove singularity in single-particle density of states has also been observed in the "momentum-selective" metal-insulator transition scenario~\cite{Gull09,Philipp09,Lin10,Gull10}, however, the quantum critical phenomena associated with the van Hove singularity and the Lifshitz transition have not been discussed there.

\begin{figure}[ht!]
\centering{}
\includegraphics[width=3.3in]{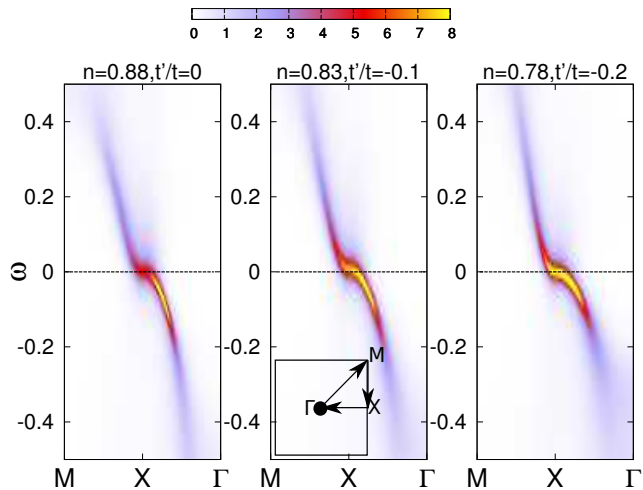}
\caption{(color online) Intensity plots of the spectral function $A(\bk,\omega)$ along the segment $M \rightarrow X \rightarrow \Gamma$ 
inside the Brillouin zone. For different $t'/t$ near the quantum critical fillings: $t'/t=0$ with $n=0.88$ (left), $t'/t=-0.1$ with $n=0.83$ (middle), and $t'/t=-0.2$ with $n=0.78$ (right), a saddle point region (flat dispersion) around $X$ crosses the Fermi level at the critical filling $n_c$. The region becomes wider as $t'$ goes to negative values.}
\label{fig:AkwSpectra_GMX}
\end{figure}

The van Hove singularity originates from the flat dispersion along the antinodal direction with energy close to the Fermi level~\cite{Radtke94,Markiewicz97,Markiewicz12}, as shown in Figure \ref{fig:AkwSpectra_GMX}. At the critical doping for $t'/t=0.0$, $-0.1$, and $-0.2$, there is always a saddle point region around momentum $X=(\pi,0)$ where the dispersion is flat. In addition, we find that the flat region become wider and more pronounced as $t'$ becomes more negative. Such flat dispersion at the chemical potential contributes low energy states and results in the van Hove singularity at the Fermi energy. Hence, our observation of the Lifshitz transition at the quantum critical doping for $t'/t\le0$ is closely tied (if not in one-to-one correspondence) to the crossing of the Fermi level
by the van Hove singularity.

\begin{figure*}[ht!]
\centering{
\includegraphics[width=2.8in,height=2.8in]{FS100tp.eps}
\hspace{0.2in}
\includegraphics[width=3.8in]{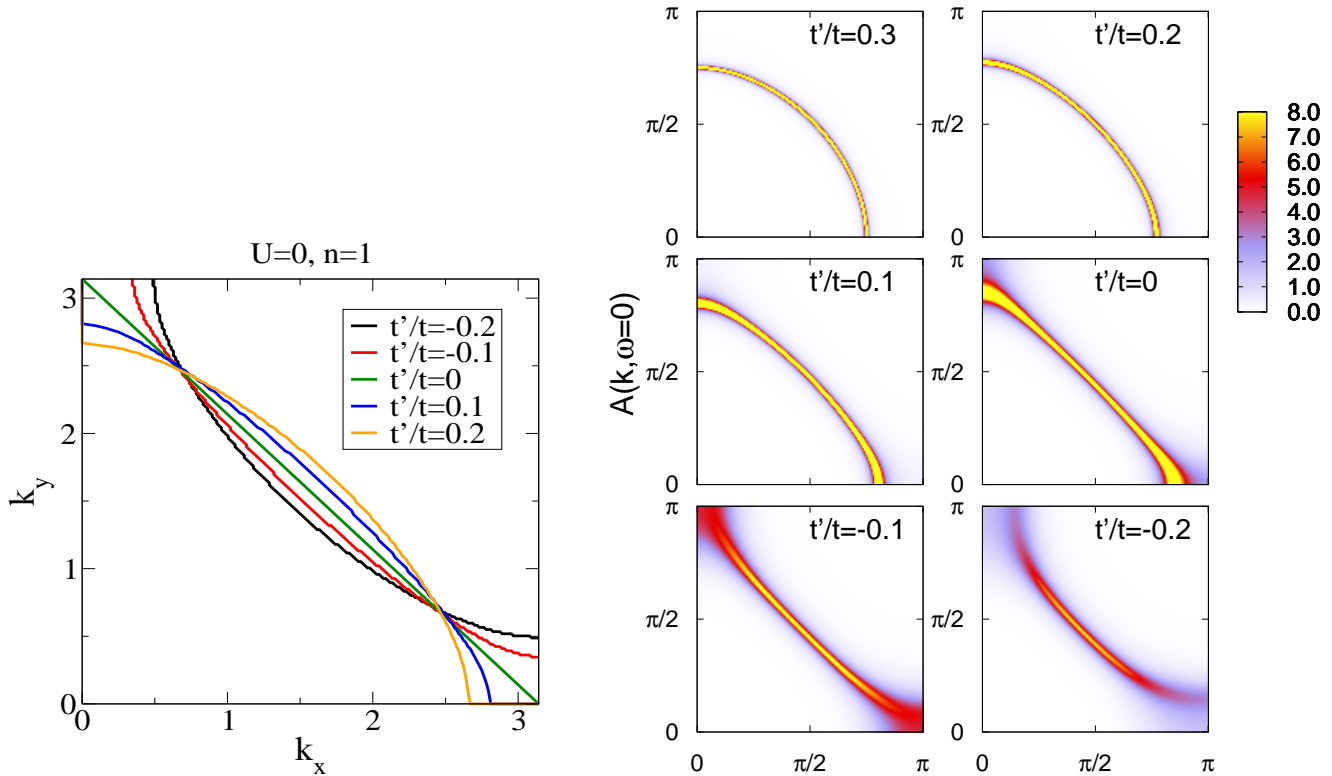}
}
\caption{(color online) $t'/t$ as the control parameter of the Lifshitz transition. (Left panel) $A(\bk,\omega=0)$ for the
non-interacting system at half-filling. As $t'/t$ goes from positive to negative, the Fermi surface changes 
from electron-like to hole-like. (Right panel) $A(\bk,\omega=0)$ for the interacting system, $U/t=6$,  with filling 
$n=0.85$ fixed. At positive $t'$, the system is inside the metallic Mott gas phase, and has complete Fermi 
surface; at negative $t'$, the system is inside the pseudogap phase, and the Fermi surface becomes hole-like. 
Since the Coulomb interaction reshuffles the spectral weight, at $t'/t=-0.1, -0.2$, the pseudogap at the antinodal 
direction can be clearly seen.}
\label{fig:DOS085_tp}
\end{figure*}

\begin{figure}
\centering{}
\includegraphics[width=3.3in]{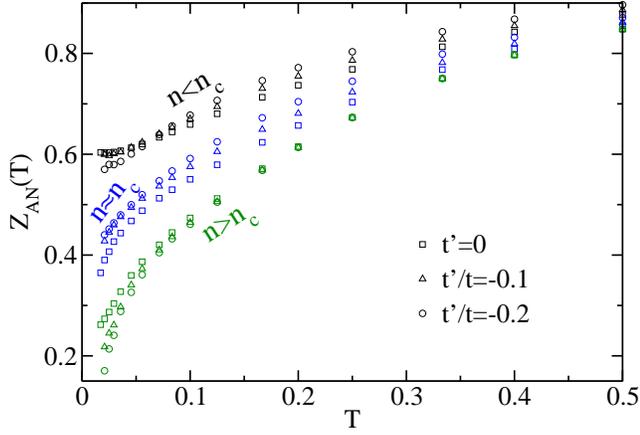}
\caption{(color online) Matsubara quasiparticle weight $Z_{AN}(T)$ versus temperature $T$ evaluated with $\bk$ on the Fermi 
surface along the antinodal direction, for various fillings and $t'$. For $n<n_c$, $Z_{AN}(T)$ displays 
Fermi liquid behavior, with $Z_{AN}(T)$ extrapolating linearly to a finite value at low $T$; for $n\approx n_c$, 
$Z_{AN}(T)$ displays the marginal Fermi liquid behavior, with negative curvature in $Z_{AN}(T)$ at low $T$; and 
once inside the pseudogap region, $n>n_c$, $Z_{AN}(T)$ goes to zero as a function of $T$ 
faster than for the marginal Fermi liquid fillings.}
\label{fig:MatZ}
\end{figure}

The next-nearest-neighbor hopping $t'$ can be viewed as the control parameter of the Lifshitz transition. To emphasize 
this point, we also calculate the zero frequency spectra for the non-interacting system as a function of $t'/t$, and 
compare them with our DCA results for the interacting system.  In Figure \ref{fig:DOS085_tp}, the left panel demonstrates 
the effect of $t'$ on the Fermi surface topology of the non-interacting half-filled system. We find the Fermi surface changes from 
electron-like to hole-like when $t'$ goes from positive to negative values. The right panel of Fig.\ \ref{fig:DOS085_tp} collects 
our results for  the spectra of the interacting system, with $n=0.85$ fixed and $t'/t$ varies from 0.3 to -0.2. Similar 
to the noninteracting system, the Fermi surface topology also changes from electron-like to hole-like as $t'/t$ goes from 
positive to negative, which confirms that $t'$ is indeed the relevant control parameter of the Lifshitz transition in 
the interacting system. However, the interacting system follows a more complicated phenomenology than the non-interacting 
system, due to the effect of electron-electron interaction. For example, in the lower two panels of the right hand side 
of Fig.~\ref{fig:DOS085_tp}, where $t'$ is negative, not only is the Fermi surface topology changed to hole-like, but the 
spectral weight along the antinodal direction $(\pi,0)$ also vanishes, signaling that the system enters the pseudogap phase.
The strong interaction results in a  redistribution of the spectral weight, leading to features different from the 
ones of the non-interacting system.

Instead of obtaining the spectral information from the analytical continued data, one can also directly read off the quasiparticle weight $Z(\bk)$ from the Matsubara frequency results. Since the quasiparticle weight will be finite across a Fermi surface, but it vanishes if the spectrum is incoherent, $Z(\bk)$ can be used to distinguish a Fermi liquid and a pseudogap state. The quasiparticle weight is calculated from the Matsubara frequency self-energy as 
$\displaystyle Z_0(\bk_F)=\big(1-\frac{\Sigma"(\bk,i\omega_0)}{\omega_0}\big)^{-1}\big|_{\bk=\bk_F}$, where $\omega_0=\pi T$ is the lowest fermionic Matsubara frequency. In the limit $T\rightarrow 0$, 
$Z_0(\bk)$ converges to the quasiparticle weight, $Z(\bk)$. Figure \ref{fig:MatZ} shows $Z_{AN}(T)=Z_0(\omega_0=\pi T, \bk_F)$, the Matsubara quasiparticle weight on the Fermi surface along the antinodal direction for various $t'/t$. $Z_{AN}$ exhibits different behavior for $n<n_c$ where the quasiparticle weight approaches a finite value and for $n \ge n_c$ where the quasiparticle weight vanishes in the limit $T\rightarrow 0$. The temperature dependence of $Z_{AN}(T)$ furthermore provides information about the relevant energy scales. One sees a clear difference in the temperature dependence of $Z_{AN}(T)$ when the system enters Fermi liquid, marginal Fermi liquid and pseudogap phases. Inside the Fermi liquid phase, $n<n_c$, where we choose $n=0.75$ for $t'/t=0$, 
$n=0.70$ for $t'/t=-0.1$ and $n=0.65$ for $t'/t=-0.2$, $Z_{AF}(T)$ extrapolates to a finite value roughly linearly at low $T$.  Close to the Lifshitz transition points, $n\approx n_c$, where we choose $n=0.88$ for $t'/t=0$, $n=0.83$ for $t'/t=-0.1$ and $n=0.78$ for $t'/t=-0.2$, $Z_{AN}(T)$ shows a behavior consistent with the marginal Fermi liquid picture, i.e.,\ a 
negative curvature in $Z_{AF}(T)$ at low $T$~\cite{Vidhyadhiraja09}.  Finally, when the system is inside the pseudogap 
phase, $n>n_c$, we choose the same filling $n=0.95$ for $t'=0, -0.1$ and $-0.2$, $Z_{AF}(T)$ goes to zero even faster 
than it does  in the marginal Fermi liquid region. The detailed information about the crossover temperatures $T^{*}$ and $T_X$, where the former signifies the crossover between pseudogap and the marginal Fermi liquid and the latter is the crossover temperature between the marginal Fermi liquid and the Fermi liquid, 
and the quantitative distinction between the marginal Fermi liquid and the pseudogap in terms 
of temperature and frequency dependence of the self energy and resistivity, have been discussed in detail 
in our previous publications~\cite{Vidhyadhiraja09,Galanakis11,Chen11}.

\begin{figure}
\centering{}
\includegraphics[width=3.3in]{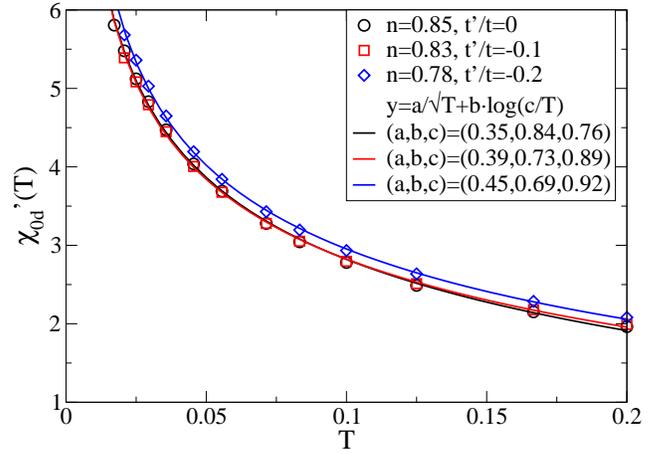}
\caption{(color online) The real part of the bare $d$-wave pairing susceptibility, $\chi'_{0d}(\omega=0,T)$, at zero frequency, close to the quantum critical fillings for $t'/t=0$, $t'/t=-0.1$, and $t'/t=-0.2$. The solid lines are fits to $\chi'_{0d}(\omega=0,T)=a/\sqrt{T}+b\log(c/T)$. Close to the quantum critical fillings ($n=0.85$ for $t'/t=0$, $n=0.83$ for $t'/t=-0.1$, and $n=0.78$ for $t'/t=-0.2$), $\chi'_{0d}(\omega=0,T)$ shows a power-law divergence with decreasing temperature. The prefactor $a$ associated with the square-root term increases as $t'$ become negative. This signifies a stronger divergence in the pairing susceptibility.}
\label{fig:ReChiw0_T_comp2}
\end{figure}

\begin{figure}
\centering{}
\includegraphics[width=3.3in]{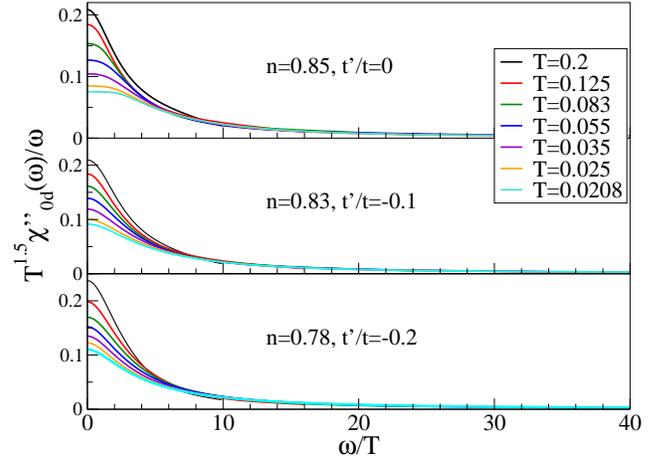}
\caption{(color online) Plots of $T^{1.5}\chi"_{0d}(\omega)/\omega$ versus $\omega/T$ close to the quantum critical doping ($n=0.85$ for $t'/t=0$, $n=0.83$ for $t'/t=-0.1$, and $n=0.78$ for $t'/t=-0.2$). As the temperature decreases, the curves coincide for $\omega/T > 9 \approx (4t/J)$ defining a scaling function $H(\omega/T)$, which corresponds to a contribution 
to $\displaystyle \chi'_{0d}(T)=\frac{1}{\pi}\int d\omega \chi"_{0d}(\omega)/\omega \propto \frac{1}{\sqrt{T}}$ as shown in Fig. \ref{fig:ReChiw0_T_comp2}. In the scaling regime, $H(\omega/T)\approx (\omega/T)^{-1.5}$.}
\label{fig:weta2_3nc}
\end{figure}

The bare $d$-wave pairing susceptibility is calculated as $\chi_{0d}(\omega)=\sum_{k}\chi_0(\omega,q=0)g_d(k)^2/\sum_{k}g_d(k)^2$, where $g_d(k)=(cos(k_x)-cos(k_y))$ is the $d$-wave form factor. It exhibits significant different features near the QCPs compared with those predicted by static van Hove scenarios~\cite{Hirsch86b,Markiewicz97}. As shown in Fig. \ref{fig:ReChiw0_T_comp2}, close to the quantum critical fillings, the real part of the bare pairing susceptibility $\chi'_{0d}(\omega=0,T)$ diverges quickly with decreasing temperature, following a power-law behavior close to $1/\sqrt{T}$, which is different from the BCS type of logarithmic divergence $\chi'_{0}(T)\propto N(0)\ln(\omega_D/T)$ with $N(0)$ the single-particle density of states at the Fermi surface and $\omega_D$ the phonon cutoff frequency. Such behavior is consistent with our previous results at $t'=0$ close to the quantum critical filling~\cite{YangSX11,Chen11}. More interestingly, 
such temperature dependence persists for $t'< 0$. 
As shown in Fig. \ref{fig:ReChiw0_T_comp2}, the prefactor, $a$, in the power-law term becomes larger as $t'$ becomes 
more negative. It signifies that the divergence becomes stronger at $t'<0$ and, if the pairing strength does not change, there will be a higher superconducting transition temperature $T_c$ compared with the one for $t'=0$. 
Moreover, when we scale the imaginary part of the bare pairing susceptibility as $T^{1.5}\chi"_{0d}(\omega)/\omega$ versus $\omega/T$~\cite{She09}, as shown in Fig. \ref{fig:weta2_3nc}, we find the curves from different temperatures fall on the same universal scaling function such that $T^{1.5}\chi"_{0d}(\omega)/\omega = H(\omega/T) \approx (\omega/T)^{-1.5}$ for $\omega/T\ge 9 \approx 4t/J$, where $J\approx0.11$~\cite{Macridin2007, Vidhyadhiraja09} is the antiferromagnetic exchange energy near half filling. From the Kramers-Kronig relation, the real and imaginary parts of the susceptibility are related via 
$\displaystyle \chi'_{0d}(T)=\frac{1}{\pi}\int d\omega \chi"_{0d}(\omega)/\omega$, so collapse of the $\chi"_{0d}(\omega)/\omega$ will contributes a term $\displaystyle \propto \frac{1}{\sqrt{T}}$ in the real part of $\chi'_{0d}(T)$.

\section{Discussion}  
\label{sec:Discussion}

\begin{figure}
\centering{}
\includegraphics[width=3.4in]{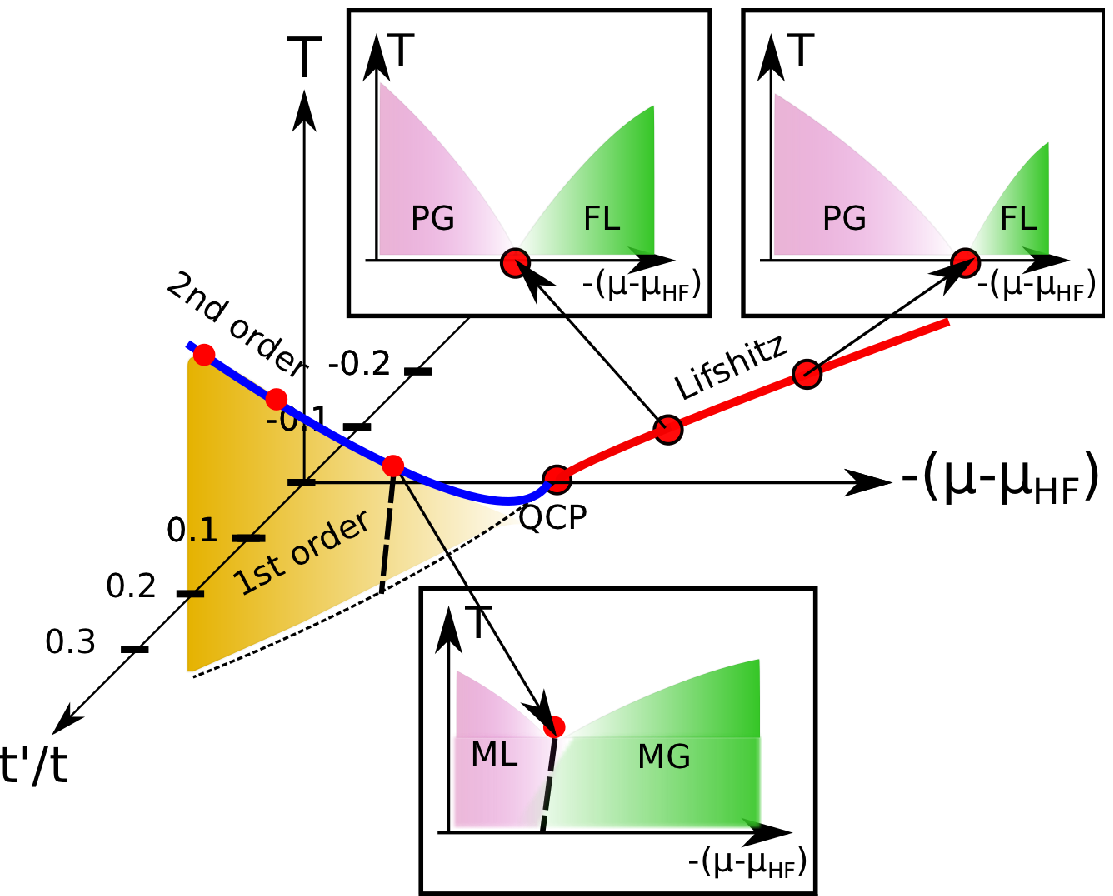}
\caption{(color online) Schematic phase diagram of the Hubbard model close to the quantum critical point (QCP) with temperature 
($T$), chemical potential ($\mu$), and next-nearest-neighbor hopping ($t'$) as the control parameters. For clarity we neglect the superconducting and antiferromagnetic phases and only focus on the Lifshitz physics. For each $t'$, we shift the chemical potential $\mu$ with respect to its half filling value $\mu_{\text{HF}}(t'/t)$. The three insets show the cut of the phase diagram at three different $t'/t=0.1,-0.1$ and $-0.2$. At positive $t'$, the Mott 
liquid and Mott gas phases are separated by a first order line; at negative $t'$, the pseudogap and the Fermi liquid phases are separated by a quantum critical region, and the region becomes wider as $t'$ moves 
towards the negative direction. As the Lifshitz points move to larger hole doping at negative $t'$, the slope of the boundary between the pseudogap and the quantum critical region becomes less steep.}
\label{fig:PhaseDiagram}
\end{figure}

One of the important issues in the field of cuprate physics is the identification of the quantum critical point accepted 
by and large as relevant for the superconductivity and in particular for the strange ``non-Fermi liquid''  behavior observed 
in the vicinity of the maximum of the superconducting dome. Based on our numerical results presented here together with 
those published previously~\cite{Khatami10,Vidhyadhiraja09,YangSX11,Galanakis11,Chen11}, we suggest the schematic phase 
diagram for the two dimensional Hubbard model shown in Figure \ref{fig:PhaseDiagram} (note that for clarity we purposely neglect the 
superconducting and the antiferromagnetic transitions in this schematic plot).  The control parameters are the next-nearest-neighbor 
hopping $t'/t$, chemical potential $\mu$, and temperature $T$. For each $t'$, we shift the chemical $\mu$ with respect to its value at half filling $\mu_{\text{HF}}(t'/t)$. For $t'>0$, there is a first order phase separation transition between an 
insulating and incompressible Mott liquid (ML region in the lower inset of Fig.~\ref{fig:PhaseDiagram}) and a weakly compressible 
metallic Mott gas (MG region in the lower inset of Fig.~\ref{fig:PhaseDiagram}). As a function of temperature, this region 
of first order transitions  terminates in a line of second order classical critical points (red dots and blue line for 
$t'>0$). By varying $t'$ this line of second order transitions is suppressed to $T=0$.  The corresponding 
$t'=0$ doping defines the quantum critical point in the phase diagram.
This QCP  separates the pseudogap regime from the Fermi liquid. 
As usual, one observes a V-shaped quantum critical region above this QCP, which shows the signatures 
of a marginal Fermi liquid. Note that this QCP at $t'=0$ also involves a Lifshitz transition, i.e.\ a change in 
Fermi surface topology from  electron-like in the Fermi liquid phase to hole-like in the pseudogap phase.
This property can be directly inferred from the changes in the low-frequency structures of the spectral function 
(see Fig.\ \ref{fig:FalseColortp0tpm01}).  At the critical doping the van Hove singularity in the single particle 
density of states is located at the Fermi energy (see Fig.\ \ref{fig:DOS}). Note that there are actually
two effects involved. Firstly, the change in the overall shape of the Fermi surface, and secondly 
a distinctive suppression of spectral weight along the antinodal directions in the pseudogap phase. Based on the momentum
resolution presently available,  we are not able to make decisive statements about a possible formation of hole pockets. 
Higher momentum resolution would be necessary to unambiguously establish the nature of the Lifshitz transition.  We are currently 
developing a multiscale dual Fermion dynamical cluster approach~\cite{YangSXDF}, which combines the DCA method used in 
this work with the recently introduced dual-fermion formalism~\cite{Rubtsov09}. This approach can systematically incorporate 
the long-ranged correlations through the dual fermion lattice calculation, and will eventually provide us the necessary 
momentum resolution to address the question of the hole pockets and the type of Lifshitz transition associated with 
the QCP. However, the numerical data presented in this work are very indicative and indeed consistent with a collapse of 
the Fermi surface from a large to a small one.

As $t'$ becomes negative phase separation is suppressed, but the critical behavior associated
with the Lifshitz transition remains, leading to a line of critical points at $T=0$, as sketched in the 
schematic phase diagram in the region $t'<0$. Since the effect of negative $t'$ is to distort the Fermi surface 
towards hole-like (see Fig.\ \ref{fig:DOS085_tp}), in the $t'/t<0$ region of the phase diagram, one necessarily needs 
higher hole doping concentration to have the van Hove singularity locate at the Fermi level. This provides us a 
simple explanation for the fact that the Lifshitz points move to higher hole doping at $t'<0$. Across the Lifshitz 
points, it is not only the change in shape of the Fermi surface, but more
importantly the strong suppression of weight along the antinodal direction which leads to the interpretation
of a Lifshitz transition at $T=0$.
The quantum fluctuations here again lead to a 
region with quantum critical behavior.
As it is commonly accepted, the low-energy model for the high-$T_c$ cuprates is the two dimensional Hubbard model with at least $t>0$ 
and $t'<0$, we thus have strong numerical evidence that the QCP is indeed due to a transition of the Fermi surface
topology, as already suggested by several other authors~\cite{Sakai09,Imada2011}. We also find at negative $t'$ that the 
quantum critical region becomes wider in the doping range (see the two upper insets of Fig.~\ref{fig:PhaseDiagram}), 
the quantum critical points move to higher hole doping concentrations, and the boundary between the pseudogap region and 
quantum critical region becomes less steep as $t'$ moves towards the negative direction.

There is, however, an important missing link. The explanation, for the interacting electron system, of why the van 
Hove singularity crosses the Fermi surface at the critical filling $n_c$, even for $t'=0$, thus triggering the Lifshitz 
transition.  In previous DCA studies of the 2D Hubbard model\cite{t_maier_02c,Gull10}, it was found that the non-local
correlations present in the DCA, but missing in the 
single site dynamic mean field approximation (DMFA), tend to distort the Fermi surface of the hole doped system so 
that it is centered around the wavevector ($\pi$,$\pi$) rather than ($0$,$0$).  As the system is doped away from half 
filling, these correlations become weaker, and the DCA Fermi surface returns to the DMFA  Fermi surface 
centered around ($0$,$0$). Thus the topology of the Fermi surface changes from hole-like to electron-like at finite 
doping.  
%
%

On the other hand, since the correlation effects at finite doping show strong temperature dependence, we expect the 
phenomena associated with the van Hove singularity also have a temperature dependence which is quite different from 
those predicted by models with a static van Hove singularity resulting from a saddle point in the bare dispersion~\cite{Markiewicz97}. 
This has important consequences for the superconducting transition $T_c$, which, at sufficiently high doping, is 
determined by the BCS-like condition $V_d\chi'_{0d}(\omega=0)=1$ where $\chi'_{0d}$ is the real part of the 
bare d-wave pairing susceptibility and $V_d$ is the strength of the d-wave pairing interaction.  

Following the results in Fig. \ref{fig:ReChiw0_T_comp2} and \ref{fig:weta2_3nc}, we find at $t'=0$, $t'/t=-0.1$, and $t'/t=-0.2$, $\chi'_{0d}(\omega=0)$ close to the van Hove singularity diverges quickly with decreasing temperature, roughly following 
the power-law behavior $\propto a/\sqrt{T}$, which is significantly different from the conventional BCS logarithmic 
divergence or the log-squared divergence found at a static van Hove singularity~\cite{Hirsch86b,Markiewicz97}. Moreover, we find that the prefactor $a$ increases with negative $t'$, which 
signifies a stronger divergence in the pairing bubble compared with the one at $t'=0$. However, we also know the $d$-wave 
pairing vertex, $V_d$, decays monotonically with hole doping, since $V_d$ originates predominantly from the spin 
channel~\cite{Maier06,YangSX11}. A negative $t'$ will frustrate the antiferromagnetic background, and hence 
suppress the spin channel contribution to the $d$-wave pairing vertex. At the same time, according to our 
Lifshitz phase diagram (Fig.~\ref{fig:PhaseDiagram}), one needs higher hole doping concentration to approach the QCP 
for negative $t'$. The higher quantum critical doping and the suppression of the spin channel contribution to the 
pairing vertex  lead us to expect the d-wave pairing vertex, $V_d$, becomes even smaller for $t'<0$ 
than it is for $t'=0$.  Hence, a more strongly divergent pairing bubble, $\chi'_{0d}$, and a  weaker pairing 
vertex, $V_d$, bring us to an interesting situation for $t'<0$, where there is a competition between $\chi'_{0d}$ and 
$V_d$. Whether this competition will yield a higher $T_c$ and larger superconducting dome in the doping range close to 
the Lifshitz points at negative $t'$ will be the subject of  future investigations. 

Our phase diagram is also phenomenologically consistent with the recent proposal of unconventional 
quantum criticality at the border of a first order metal-insulator transition and a continuous transition 
triggered by quantum fluctuations~\cite{Yamaji06,Misawa07,Yamaji2007}. This marginal quantum critical point 
belongs to an unprecedented universality class and has unique feature with combined characteristics of symmetry breaking and 
topological (Lifshitz) transitions~\cite{Imada2010}. Whether the Lifshitz points presented at $t'/t\le0$ in this work 
display unconventional quantum criticality is also a subject of further studies.

\section{Conclusion}
\label{sec:Conclusion}
Using large-scale dynamical cluster quantum Monte Carlo simulations, we map out the Lifshitz phase diagram of the two dimensional Hubbard model in the vicinity of the quantum critical fillings. The control parameters of the 
phase diagram are temperature $T$, chemical potential, and the next-nearest-neighbor hopping $t'/t$. Consistent with 
our previous results~\cite{Khatami10,Chen11,Galanakis11}, we find at  positive $t'$ a first order phase 
separation transition which is terminated by a second order critical point. As $t'\rightarrow 0$, the second order 
terminus is driven to zero temperature, and becomes the QCP separating the pseudogap and the Fermi liquid phases. 
Here, we extend the investigation into negative $t'$, and find out 
a line of van Hove singularities where the Fermi surface topology changes from electron-like in the Fermi liquid 
region to hole-like in the pseudogap region of the phase diagram. The points on this line of van Hove singularities 
hence are the quantum critical points where the Lifshitz transition occurs. Close to these QCPs, the bare $d$-wave pairing 
bubble diverges algebraically in temperature. Originating from these Lifshitz points, the V-shaped quantum critical 
region emerges with marginal Fermi liquid properties and vanishing quasiparticle weight~\cite{Vidhyadhiraja09,Chen11}. 
We also find the V-shaped quantum critical region becomes wider in doping range (and chemical potential) as $t'/t$ 
becomes negative.

There remain a number of interesting open issues, including the possible formation of hole pockets in the pseudogap 
phase, the reason of the van Hove singularity crossing the Fermi surface at the quantum critical doping, the superconducting 
transition temperature $T_c$ and the shape of the superconducting dome at negative $t'$, and the possible unconventional 
quantum criticality associated with the Lifshitz transition points. All these questions require not only massively parallel 
large-scale simulations, but more importantly, new techniques that will greatly increase the momentum resolution. In fact,
 progress has already been made along these directions~\cite{YangSXDF}, and the remaining questions will be addressed in 
future work.

\begin{acknowledgments}
We would like to thank Ka-Ming Tam, Shu-Xiang Yang, Sandeep Pathak, Karlis Mikelsons, Jian-Huang She, and J. Zaanen 
for useful conversations. We thank Kai Sun and E. Gull for valuable comments. We also thank Carol Duran for careful 
reading of the manuscript. This work is supported by NSF 
grants OISE-0952300 and DMR-0706379, and the NSF EPSCoR Cooperative Agreement No. EPS-1003897 with additional support 
from the Louisiana Board of Regents. TP acknowledges the support by the DFG through the research unit FOR 1346 and the 
DAAD through the PPP exchange program. Supercomputer support was provided by the NSF XSEDE under grant number 
DMR100007.
\end{acknowledgments}

\bibliography{Lifshitz}

\begin{thebibliography}{57}
\bibitem[{foo()}]{footnote12}
\bibinfo{note}{The cumulant interpolation scheme yields the hole pocket in the
  underdope region~\cite{Sakai09,Sakai10,Sakai12}, the corresponding Lifshitz
  transition from Fermi liquid to pseudogap will be a "pocket-vanishing" type
  different from our findings of "neck-collapsing" type with self energy
  interpolation scheme. But the two schemes yield transitions at the same
  critical filling $n_c$.}
\end{thebibliography}

\end{document}